\documentclass[11pt]{article}
\pdfoutput=1
\usepackage{amssymb,amsmath,amsthm,graphicx,ulem}
\usepackage{graphicx}
\usepackage{epsfig}
\usepackage{bm} 
\usepackage{amsfonts,amssymb,amsmath}
\usepackage{latexsym}
\usepackage[english]{babel}
\usepackage{subfigure}

\def\beq{\begin{equation}}
\def\eeq{\end{equation}}
\def\bea{\begin{eqnarray}}
\def\eea{\end{eqnarray}}

\newcommand{\eqn}{\begin{eqnarray}}
\newcommand{\eeqn}{\end{eqnarray}}
\newcommand{\arr}{\begin{eqnarray*}}
\newcommand{\earr}{\end{eqnarray*}}

\newcommand{\eg}{{e.g.}\ }
\newcommand{\ie}{{i.e.}\ } 

\newcommand{\lp}{\left(}
\newcommand{\rp}{\right)}

\def\ii{\textrm{i}}

\begin{document}

\setlength{\unitlength}{1mm}

\thispagestyle{empty} 
 \vspace*{2cm}

\begin{center}
{\bf \LARGE  Ultraspinning instability of anti-de Sitter black holes}\\

\vspace*{2.5cm}

{\bf \'Oscar J.~C.~Dias$^{a}\,$}, {\bf Pau Figueras$^{a}\,$}, \\
{\bf Ricardo Monteiro$^{b}\,$}, {\bf Jorge E.~Santos$^{c}\,$}
\vspace*{0.5cm}

{\it $^a\,$ DAMTP, Centre for Mathematical Sciences, University of Cambridge,\\
Wilberforce Road, Cambridge CB3 0WA, United Kingdom}\\[.3em]
{\it $^b\,$ The Niels Bohr International Academy, The Niels Bohr Institute, \\
Blegdamsvej 17, DK-2100 Copenhagen, Denmark}\\[.3em]
{\it $^c\,$ Department of Physics, UCSB, Santa Barbara, CA 93106, USA}\\[.3em]

\vspace*{0.5cm} {\tt O.Dias@damtp.cam.ac.uk, p.figueras@damtp.cam.ac.uk, \\
monteiro@nbi.dk, jss55@physics.ucsb.edu \\ \phantom{a}}

\end{center}

\begin{abstract}

Myers-Perry black holes with a single spin in $d>5$ have been shown to be unstable if rotating sufficiently rapidly. We extend the numerical analysis which allowed for that result to the asymptotically AdS case. We determine numerically the stationary perturbations that mark the onset of the instabilities for the modes that preserve the rotational symmetries of the background. The parameter space of solutions is thoroughly analysed, and the onset of the instabilities is obtained as a function of the cosmological constant. Each of these perturbations  has been conjectured to represent a bifurcation point to a new phase of stationary AdS black holes, and this is consistent with our results.

\end{abstract}

\noindent

\vfill \setcounter{page}{0} \setcounter{footnote}{0}
\newpage



\setcounter{equation}{0}
\section{Introduction}

There has been recent progress in understanding the phase diagram of
higher-dimensional  asymptotically flat vacuum black holes. The
purpose of this work is to extend some of the techniques used to the
case of asymptotically anti-de Sitter (AdS) black holes.

Let us first review the asymptotically flat case. While the
four-dimensional Kerr  black hole is unique and (expected to be)
stable, the higher-dimensional picture is much richer
\cite{Emparan:2008eg}. In $d=5$, the discovery of the black ring by
Emparan and Reall \cite{Emparan:2001wn} showed that it can have the
same conserved charges as the Myers-Perry (MP) black hole
\cite{myersperry}. This non-uniqueness discovery triggered the
research in higher dimensional black holes and  a variety of
explicit rotating solutions have been recently found: black Saturns
\cite{Elvang:2007rd}, concentric rings
\cite{Iguchi:2007is,Evslin:2007fv}, orthogonal rings
\cite{Izumi:2007qx, Elvang:2007hs}, and generalisations thereof.
Consider solutions rotating on a single plane which are in thermal
equilibrium (i.e. in the case that there are disconnected components
of the event horizon, these have the same temperature and angular
velocity). One interesting fact is that in the zero temperature
limit  the new solutions {\it and} the MP black hole coincide in the
same nakedly singular solution \cite{Elvang:2007hg}. In $d>5$, this
limit does not exist, and the MP black hole has an unbounded angular
momentum for a given mass. In a certain sense, the singular limit is
`resolved'. However, does the MP black hole still connect to the new
solutions? In $d>5$, the black ring and the other solutions have not
been constructed exactly, but are expected to exist. In fact,
approximate methods for solving the Einstein equations in the
ultraspinning limit (large angular momentum with respect to the mass
scale) indicate that they do exist
\cite{Emparan:2007wm,Emparan:2009cs,Emparan:2009at,Emparan:2009vd}.

The answer seems to be that the MP black hole connects to the other
solutions through yet more families of black holes. These families
bifurcate from the MP branch at the (stationary) onset of the so
called  `ultraspinning instabilities'. The first step in
understanding this connection was given by Emparan and Myers
\cite{Emparan:2003sy}, who argued that the singly-spinning (i.e.
rotating on a single plane) MP black hole should be unstable if
rotating sufficiently rapidly. The reasoning is that if the angular
momentum per unit mass is large enough, MP black holes start
behaving like black branes, since their horizons become disk-like
along the rotation plane in this ultraspinning limit. But it has
been shown by Gregory and Laflamme \cite{Gregory:1993vy} that black
branes are unstable and therefore one concludes that MP black holes
should be unstable for a sufficiently large (but finite) angular
momentum per unit mass. Emparan and Myers further pointed out that
at the onset of such an instability,  there should exist a
stationary perturbation that preserves the rotational symmetries of
the background.  This perturbation would then signal the existence
of  a new family of black holes with spherical horizon topology.
This is analogous to what happens in the black brane case: the
threshold mode of the Gregory-Laflamme instability is the
perturbative signal for a family of non-uniform branes
\cite{Gubser:2001ac,Wiseman:2002zc}. Ref.~\cite{Emparan:2007wm}
conjectured that the new families bifurcating from the MP branch,
when continued  along the phase diagram, would connect continuously
to the black ring branch,  the black Saturn branch, and so on.
Therefore, the connection of non-uniqueness with instabilities
provides a partial understanding of the proliferation of higher
dimensional black hole phases.

Recently, these conjectures were put on a firmer footing when it was
shown numerically that the singly-spinning MP black hole does indeed
possess such threshold modes \cite{Dias:2009iu,Dias:2010maa}.
Furthermore, these perturbations exhibit an underlying harmonic
structure and induce deformations on the shape of the horizon which
are consistent with the proposal that the new families will connect
to the black ring, to the black Saturn, etc. Ref.~\cite{Dias:2009iu}
also conjectured that these classical instabilities can only occur
when the black hole possesses at least two distinct local
thermodynamic instabilities, as we shall review later. This is a
necessary but not sufficient condition. The reason is that local
thermodynamic instabilities are associated to the lowest harmonics,
and thus with the asymptotic charges, the mass and the angular
momentum. Higher harmonics, which cannot change the asymptotic
charges and are associated to the bifurcation to new families,
should only become unstable for rotations higher than the
thermodynamic modes. This conjecture allows for any number of
independent angular momenta, and has already been verified in the
instability of $d>5$ cohomogeneity-1 MP solutions (equal independent
spins in odd $d$) \cite{Dias:2010eu}. This sector of MP black holes
has a regular extremal limit and the angular momentum is bounded
from above, yet the conjecture of Ref.~\cite{Dias:2009iu} indicates
that an instability is possible. Indeed linear perturbations growing
exponentially with time were found, which is an important check
since the previous work could only determine the stationary
threshold modes.

In this paper, we will extend the results in
Refs.~\cite{Dias:2009iu,Dias:2010maa} to the asymptotically AdS
case. The properties of asymptotically AdS spacetimes have been
greatly explored due to the AdS/CFT correspondence, which equates
quantum gravity (string theory) in an asymptotically AdS spacetime
to a conformal field theory (CFT) living on the boundary of that
spacetime \cite{Maldacena:1997re,Aharony:1999ti}. In particular, the
phase diagram of black holes in AdS is in direct correspondence to
the phases of the dual CFT at finite temperature. Therefore, even if
many higher-dimensional black hole solutions present classical
instabilities, or do not dominate thermodinamically the
gravitational partition function, they still provide a valuable
insight into the CFT phases.

\begin{figure}[t] \centering
\includegraphics[width = 7 cm]{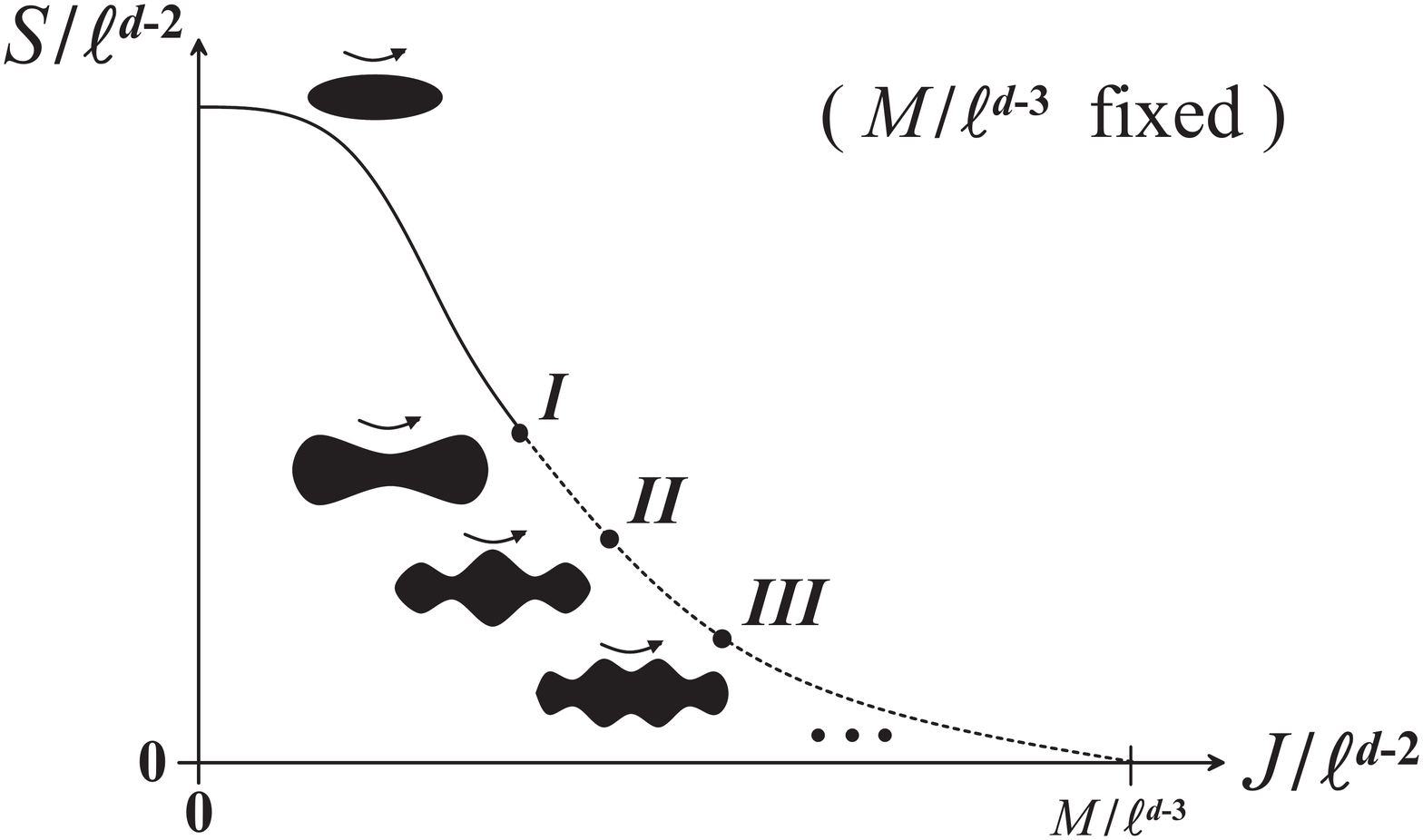}
\caption{\label{fig:phases}Phase diagram of singly-spinning MP-AdS
black holes in $d\geq 6$. We plot the entropy $S$ vs.\ the angular
momentum $J$, at a fixed value of the mass $M$, in units of the AdS
curvature radius $\ell$. The figure illustrates the conjecture of
Ref.~\cite{Caldarelli:2008pz}. At sufficiently large spin the MP-AdS
solution becomes unstable for axisymmetric perturbations (dashed
line), and at the threshold of the instability a new branch of black
holes with a central pinch appear ($I$). As the spin grows, new
branches of black holes with further axisymmetric pinches ($II$,
$III$, \ldots) appear. We determine numerically the points where the
new branches appear, but it is not yet known in which directions
they run.}
\end{figure}

Black rings and black Saturns in AdS have been constructed in
certain  approximations \cite{Caldarelli:2008pz}\footnote{See
Ref.~\cite{AdSblackfolds} for the extension of the more systematic
{\it blackfold} approach of
Refs.~\cite{Emparan:2009cs,Emparan:2009at,Emparan:2009vd} to
asymptotically AdS black holes.}, and the conjectures of
Ref.~\cite{Emparan:2007wm} regarding the connections between
singly-spinning black hole families were extended to the
asymptotically AdS case also in Ref.~\cite{Caldarelli:2008pz}.
Notice that singly-spinning MP-AdS black holes \cite{Hawking:1998kw}
(and indeed any asymptotically AdS stationary black hole
\cite{Chrusciel:2006zs}) present a BPS-type upper bound on their
angular momentum, $|J|<M\ell$ (with $M$ being the mass and $\ell$
the cosmological length), while there is no such bound in the
asymptotically flat case. However, the instabilities/bifurcations
are still expected as we increase the rotation up to that bound
\cite{Caldarelli:2008pz} (see Fig.~\ref{fig:phases}).
In this paper we will numerically determine the dependence of the
stationary  threshold modes on the cosmological constant. The
procedure is an extension to $d \geq 6$ of the $d=4$ Kerr-AdS
analysis \cite{Monteiro:2009ke}, where no classical instability was
found; the latter was itself an extension of the Schwarzschild-AdS
case \cite{Prestidge:1999uq}.

We confirm the conjecture of Ref.~\cite{Dias:2009iu}, which gives a
necessary  but not sufficient condition for the onset of the
ultraspinning instability. In this paper we generalise the previous
formulation  so that it can be applied to  the asymptotically AdS
case. We find that all bifurcations occur when the MP-AdS black hole
is unstable under superradiance
\cite{Cardoso:2004hs,Kunduri:2006qa,Cardoso:2006wa}, which implies
that the new families of black holes will inherit this instability.

Before proceeding, let us mention that asymptotically flat
singly-spinning MP  black holes also suffer from an instability that
breaks their axisymmetry \cite{Shibata:2009ad,Shibata:2010wz}. This
occurs even in $d=5$, and for $d \geq 6$ it sets in for lower
rotations than the ultraspinning instability that we  study here.
However, the threshold mode is not associated to new stationary
black holes. It would be interesting to find whether this
non-axisymmetric instability also extends to asymptotically AdS
black holes and to understand how it relates to the AdS superradiant
instability.

This paper is organised as follows. In section~\ref{sec:ultraspin},
we review the properties  of singly-spinning MP-AdS black holes and
discuss the ultraspinning instability. In
section~\ref{sec:PerturbationProblem}, we detail the linear
perturbations problem that is solved numerically. We conclude in
section~\ref{sec:results}, with the discussion of the results.

\setcounter{equation}{0}
\section{Myers-Perry-AdS black holes}
\label{sec:ultraspin}

In this section, we will start by reviewing the properties of
singly-spinning  MP-AdS black holes. Although for such solution the
angular momentum (at fixed mass) has an upper bound, we provide a
thermodynamic argument, discussed in
subsection~\ref{subsec:ultrathermo}, according to which these black
holes {\it may} be afflicted by the ultraspinning instability. This
motivates the numerical work, described in later sections, of
searching for the onset of the instability. We conclude this section
by noting that all the AdS black holes potentially afflicted from
the ultraspinning instability are also afflicted by the AdS
superradiant instability.

\subsection{Solution}
\label{subsec:MPads}

The four-dimensional Kerr-AdS black hole was found by Carter
\cite{Carter:1968ks},  and its extension to higher dimensions, when
rotation on a single plane is considered, was obtained by Hawking,
Hunter and Taylor \cite{Hawking:1998kw}.\footnote{Since the
Hawking-Hunter-Taylor solution usually refers to the
five-dimensional black hole with two independent spins, also found
in \cite{Hawking:1998kw}, we will refer to the black hole
represented by \eqref{mpbh} simply as (singly-spinning)
Myers-Perry-AdS black hole.} The metric is given by
\begin{equation*}
\begin{aligned}
ds^2 =& -\frac{\Delta_r}{\Sigma}\left( dt-\frac{a\sin^2\theta}{\Xi}
\,d\phi \right)^2 + \frac{\sin^2\theta\,\Delta_\theta}{\Sigma}\left[
\frac{r^2+a^2}{\Xi}\,d\phi-a\,dt\right]^2
+\frac{\Sigma}{\Delta_r}\,dr^2+
\frac{\Sigma}{\Delta_\theta}\,d\theta^2 \\
& +r^2\cos^2\theta\, d\Omega^2_{(d-4)}\,,\label{mpbh}
\end{aligned}
\end{equation*}
where $d\Omega^2_{(d-4)}$ is the line element of a unit-radius $(d-4)$-sphere and
\begin{equation}\label{mpbh:aux}
\Delta_r=(r^2+a^2)\lp
1+\frac{r^2}{\ell^2}\rp-\frac{r_m^{d-3}}{r^{d-5}}\,,\qquad
\Delta_\theta=r^2+a^2\cos^2\theta\,,\qquad
\Sigma=r^2+a^2\cos^2\theta\,,\qquad \Xi=1-\frac{a^2}{\ell^2}\,.
\end{equation}
This solution of the Einstein equations with negative cosmological
constant,  $R_{\mu\nu} = -(d-1)\ell^{-2}\, g_{\mu\nu}$, is
parameterised by three length scales, namely the AdS curvature
radius $\ell$, the mass-radius $r_m$ and the rotation parameter $a$.
The Komar mass $M$ and angular momentum $J$ of the black hole are
given in terms of these parameters by \cite{Gibbons:2004ai} \beq
\label{MP:MJ} M=\frac{ {\cal A}_{d-2} r_m^{d-3}}{8\pi G\,\Xi^2}\lp
1+\frac{d-4}{2}\,\Xi\rp\,, \qquad J=\frac{{\cal A}_{d-2}}{8\pi G}
\frac{ r_m^{d-3}}{\Xi^2}\,a \,, \eeq where ${\cal A}_{d-2} =
2\,\pi^{(d-1)/2} / \Gamma[(d-1)/2]$ is the volume of a unit-radius
$(d-2)$-sphere, and $G$ denotes Newton's constant which we set to
one ($G=1$). The solution satisfies the BPS-like bound
\cite{Chrusciel:2006zs}
\begin{equation}
|J|<M\ell \qquad \Leftrightarrow \qquad |a|<\ell \,.
\label{MP:constrainta}
\end{equation}
The metric \eqref{mpbh} does not describe a black hole if this bound
is saturated.  In the limit $|a| \to \ell$ either the charges (mass
and angular momentum) and the horizon equator circumference of the
black hole diverge, or the horizon vanishes and the solution is
nakedly singular.

The event horizon lies at the largest real root $r=r_+$ of
$\Delta_r(r)=0$.  The horizon angular velocity measured with respect
to a non-rotating frame at infinity is
\cite{Gibbons:2004ai,Caldarelli:1999xj}
\begin{equation} \label{MP:OmegaH}
\Omega_H=\frac{a}{r_+^2+a^2}\left ( 1+\frac{r_+^2}{\ell^2} \right
)\,,
\end{equation}
while the temperature and entropy of the black hole are
\cite{Gibbons:2004ai}
\begin{eqnarray} \label{MP:TempEnt}
 && T_H=\frac{1}{4\pi \,r_+(a^2+r_+^2)} \left( r_+^2 \left[d-3 + (d-1)\frac{r_+^2}{\ell^2} \right]
 + a^2 \left[d-5 + (d-3)\frac{r_+^2}{\ell^2} \right] \right) \,,\nonumber\\
 && S=\frac{{\cal A}_{d-2}}{4}\frac{(r_+^2+a^2)r_+^{d-4}}{\Xi}\,.
\end{eqnarray}

In $d=4$, the temperature vanishes for $|a|= r_+ \sqrt{(3
r_+^2+\ell^2)(\ell^2-r_+^2)^{-1}}$.  However, as a consequence of
the constraint \eqref{MP:constrainta}, we have an extremal regular
(\ie with finite size horizon area)  black hole if and only if
$r_+/\ell<3^{-1/2}$. In $d=5$, like in the asymptotically flat space
limit $\ell\rightarrow \infty$, the temperature goes to zero at the
singular limit $|a|\rightarrow r_m$, whereas the bound on the
rotation parameter for a finite $r_+$ is \eqref{MP:constrainta}. For
$d\geq 6$, the temperature can never vanish and thus there is no
extremal solution. The parameter space is bounded only by
\eqref{MP:constrainta}. This is in sharp contrast with the
asymptotically flat case, in which MP black holes rotating on a
single plane in $d\geq 6$ have no upper bound on their angular
momentum.

\subsection{Thermodynamic zero-modes and the ultraspinning regime \label{subsec:ultrathermo}}

The condition of local thermodynamic stability for a black hole with charges $M,J_i$ and entropy $S$ is the positivity of the Hessian
\begin{equation}
\label{thermoHessian} -S_{\alpha\beta}\equiv
-\,\frac{\partial^2{S(x_\gamma)}}{{\partial x_\alpha}{\partial
x_\beta}} \,, \qquad \; x_\alpha=(M,J_i)\,.
\end{equation}
We write the condition in its most general form, \ie taking into
account the several   angular momenta allowed in higher dimensions,
$i=1,\ldots,\lfloor (d-1)/2 \rfloor$ (where $\lfloor\,.\,\rfloor$
stands for the smallest integer part). In this section, we will
review the arguments of Refs.~\cite{Dias:2009iu,Dias:2010eu} which
relate the eigenvalues of this {\it thermodynamic} Hessian to the
occurrence of {\it classical} instabilities, namely the
ultraspinning instability.

There are two general properties of this Hessian for AdS black
holes: (i) For all (non-extremal)  asymptotically flat vacuum black
holes, in any $d$, the Hessian has at least one negative eigenvalue
\cite{Dias:2010eu}; so, by continuity, black holes whose size is
much smaller than the AdS curvature radius should be
thermodynamically unstable. (ii) Large black holes in AdS, however,
are expected to be thermodynamically stable, and thus the Hessian
should be positive definite.

Let us take the $d=4$ Kerr-AdS case, where we have a $2\times 2$
thermodynamic Hessian.  One of the eigenvalues is positive in the
entire parameter space. However, the second eigenvalue can be
negative, null or positive depending on the parameters. The null
point is where the specific heat at constant angular velocity
diverges, changing sign. We shall say that, at the surface in
parameter space where the Hessian becomes degenerate, we have a {\it
thermodynamic zero-mode}, corresponding to the eigenvector with zero
eigenvalue. The situation is analogous in the singly-spinning $d=5$
case (say $J_2=0$), with two eigenvalues being always positive, and
one changing sign. In the $d \geq6 $ singly-spinning case (say
$J_i=0$ for $i>1$) things are more interesting. {\it Two} of the
eigenvalues of the thermodynamic Hessian change sign, and all the
others remain positive throughout the parameter space. So there will
be two thermodynamic zero-modes, marking the onsets of two distinct
local thermodynamic instabilities.

\begin{figure}[t] \centering
\includegraphics[width = 8 cm]{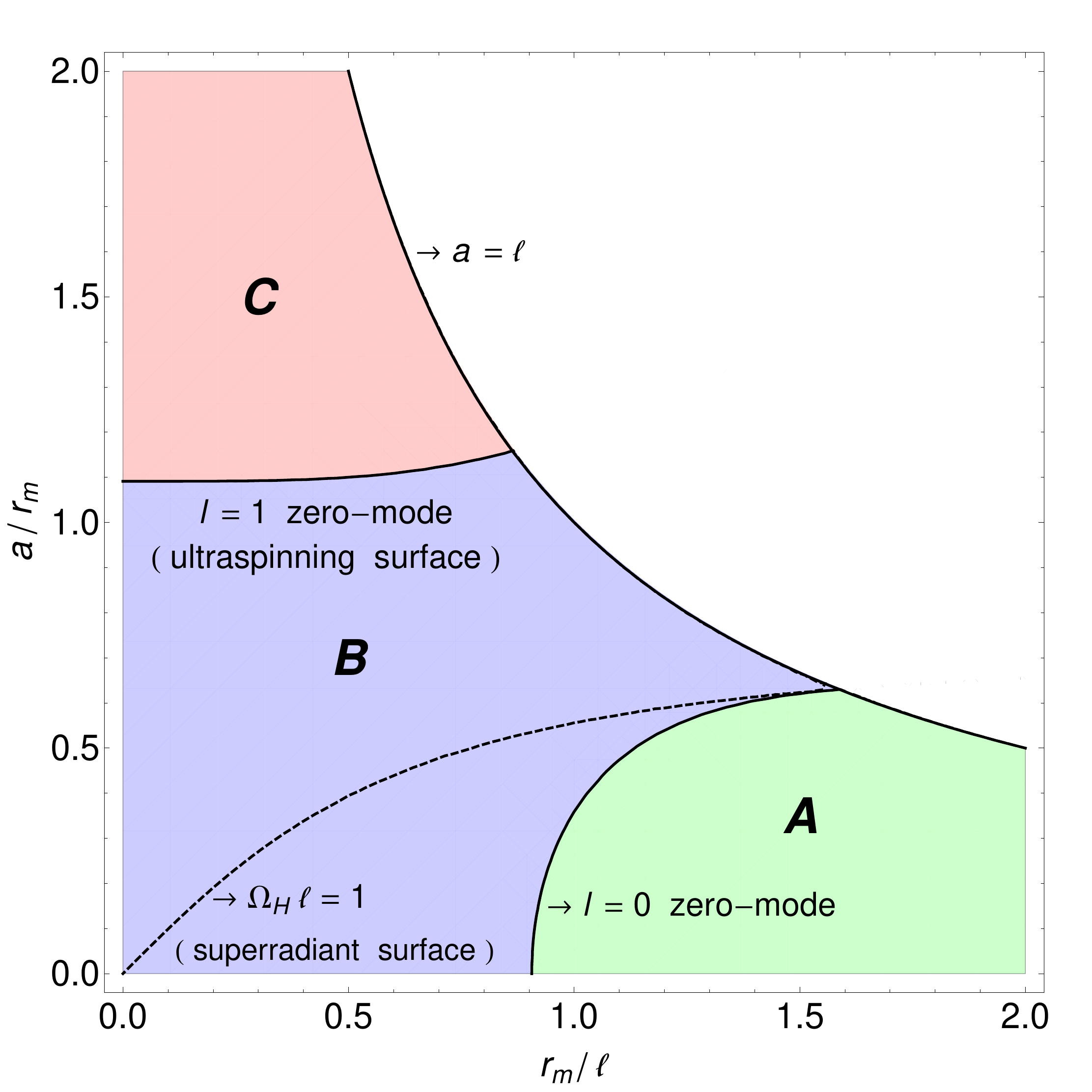}
\caption{\label{fig:thermo6d} Parameter space of singly-spinning
Myers-Perry$-$AdS black holes in $d=6$: dimensionless rotation
parameter $a/r_m$ as a function of the dimensionless mass-radius
parameter $r_m/\ell$. Regular black holes exist only for $a<\ell$.
The thermodynamic $l=0$ and $l=1$ zero-mode curves, described by
\eqref{Hessian0modeL0} and \eqref{Hessian0modeL1}, are plotted. The
thermodynamic $l=1$ zero-mode curve defines also the ultraspinning
surface, above which (region $C$) the black holes might be
ultraspinning unstable (later section~\ref{sec:results} and
Fig.~\ref{fig:spectrum} confirm this is indeed the case). The
superradiant curve $\Omega_H\ell=1$ is also plotted. Above this
curve all black holes are superradiant unstable. For $d>6$ the plot
is qualitatively very similar.}
\end{figure}

Let us consider the parameterisation $(r_+,a)$, which uniquely
specifies a singly-spinning MP-AdS black hole,  for a given AdS
curvature radius $\ell$. Recall that $|a|<\ell$ is the bound on the
parameter space, as represented in Figure~\ref{fig:thermo6d} for
$d=6$ (it is qualitatively similar in $d>6$). Now, for large
$r_+/\ell$, the black hole is thermodynamically stable. As we
decrease $r_+$, one of the eigenvalues changes sign first, which
corresponds to the transition between regions $A$ and $B$ in
Figure~\ref{fig:thermo6d}. Let us label the associated zero-mode as
the $l=0$ zero-mode, for reasons that will  become apparent later.
Its parameter space locus is:\footnote{Notice that the Hawking-Page
transition \cite{Hawking:1982dh} occurs for $r_+/\ell$ larger than
\eqref{Hessian0modeL0}. This implies that hot AdS space is preferred
(in the grand-canonical ensemble) over these black holes if
$r_+<\ell$, in any $d$. Global thermodynamic stability is stricter
than local thermodynamic stability.}
\begin{equation}
\hbox{Thermodynamic $l=0$ zero-mode ($d\geq 4$):}\quad
r_+^2=\frac{d-3}{2(d-1)}\left(a^2+\ell^2+\sqrt{a^4-\gamma_d \,a^2
\ell^2+\ell^4}\right)\,, \label{Hessian0modeL0}
\end{equation}
where $\gamma_d \equiv {2 \left(d^2-6 d+1\right)}/{(d-3)^2}$.

The $l=0$ zero-mode is present for any $d\geq 4$. However, for
$d\geq 6$ there is an additional zero-mode,  corresponding to the
second eigenvalue that becomes negative as we lower $r_+$ for fixed
$a$. This corresponds to the transition between regions $B$ and $C$
in Figure~\ref{fig:thermo6d}. Let us label the associated zero-mode
with $l=1$. Its parameter space locus is:
\begin{equation}
\hbox{Thermodynamic $l=1$ zero-mode ($d\geq 6$):}\quad
r_+^2=\frac{d-3}{2(d-1)}\left(a^2+\ell^2-\sqrt{a^4-\gamma_d \, a^2
\ell^2+\ell^4}\right)\,. \label{Hessian0modeL1}
\end{equation}

What happens as we decrease $r_+$ further for fixed $a$? As
discussed in the Introduction,  singly-spinning MP black holes were
conjectured to be unstable if rotating too rapidly, i.e. $|a| \gg
r_+$, since they start behaving like black branes
\cite{Emparan:2003sy}. In the asymptotically AdS case, there should
also be an analogous unstable regime \cite{Caldarelli:2008pz}. Now,
there can be two types of instabilities. One for which the
perturbations break the rotational symmetry of the background, i.e.
the $\partial_\phi$ Killing vector of \eqref{mpbh}; see
\cite{Shibata:2009ad,Shibata:2010wz} for the analysis of these
perturbations in the asymptotically flat case. And one for which the
spatial isometries of the background are preserved. Instabilities of
the latter type admit stationary threshold modes, which may
correspond to bifurcations to new black hole phases
\cite{Emparan:2003sy,Emparan:2007wm}. Ref.~\cite{Emparan:2003sy}
suggested that an order of magnitude estimate for the critical
rotation at the bifurcation point  could be given by what we
described above as the $l=1$ zero mode. In fact, this point marks
the transition to the black brane-like behaviour and it  only occurs
for $d\geq6$. The subsequent understanding of the problem allowed
for a more precise formulation of the conjecture, which can also be
extended beyond the singly-spinning case.

Indeed, the {\it ultraspinning conjecture} of
Ref.~\cite{Dias:2009iu} proposes that instabilities whose onset is a
stationary and axisymmetric mode, i.e. a mode which gives a
bifurcation to a new stationary black hole family, can only occur
after the $l=0$ and $l=1$ zero-modes, i.e. in region $C$ of
Figure~\ref{fig:thermo6d} in the present case. This is a necessary
but not sufficient condition. The reasoning is that the onset of
these instabilities corresponds to the $l\geq2$ zero-modes, as we
shall review below.

Ref.~\cite{Reall:2001ag}, in the static case, and
Refs.~\cite{Monteiro:2009tc,Dias:2010eu},  in the general case,
showed that a zero-mode (or a negative mode, the eigenvector of a
negative eigenvalue) of the thermodynamic Hessian
\eqref{thermoHessian} is also a stationary and axisymmetric
zero-mode (negative mode) of the black hole Euclidean action. Since
we are dealing with stationary and axisymmetric modes, this result
extends to the Lorentzian action. In the case of a zero-mode, it
corresponds to a classical perturbation of the black hole which
preserves its temperature and angular velocities, but changes its
asymptotic charges (mass and angular momenta). The change in the
asymptotic charges can simply be inferred from the eigenvector
$(\delta M, \delta J_i)$ of the Hessian \eqref{thermoHessian}
associated with the null eigenvalue. In the case of a negative mode,
it corresponds to both: (i) an off-shell perturbation of the black
hole which makes the (Euclidean) gravitational partition function
pathological, the expected signal of a local thermodynamic
instability; and (ii) an on-shell stationary perturbation of a black
brane, where the negative eigenvalue corresponds to (minus) the
`mass-squared' coming from the dimensional reduction along the brane
directions. This connection is the basis of the Gubser-Mitra
conjecture \cite{Gubser:2000ec}.

The first example of the relation between negative modes of the
action and  thermodynamic stability was the determination of the
negative mode of the Schwarzschild black hole \cite{Gross:1982cv}.
The existence of such mode is a direct consequence of the fact the
Schwarzschild black hole has negative specific heat.  Indeed, the
extension to the Schwarzschild-AdS case \cite{Prestidge:1999uq}
showed that the relevant mode changes sign exactly where the
specific heat changes sign, as expected. This result was extended to
the Kerr-AdS case \cite{Monteiro:2009ke}, where the numerical
techniques used in
\cite{Dias:2009iu,Dias:2010eu,Dias:2010maa,Dias:2010ma}, and also in
this paper, were applied first to a black hole stability problem.

Now we can justify the labels $l=0$ and $l=1$ used before, following
\cite{Dias:2009iu}.  They are based on the harmonic structure of the
metric perturbations corresponding to the zero-modes. The $l=0$
zero-mode has no nodes, i.e. the metric perturbation does not vanish
anywhere on the horizon, whereas the $l=1$ zero-mode has one node,
as plotted in \cite{Dias:2009iu} for the singly-spinning MP case.
Now, the $l=0$ and $l=1$ zero-modes are associated with the
asymptotic charges of the
spacetime.\footnote{Ref.~\cite{Dias:2010eu} analysed MP black holes
with equal spins in odd $d$, which have the convenient property of
being cohomogeneity-1. In that case, the number $l$ labels a precise
harmonic of the $CP^N$ base space, and it can be shown that the
$l=0$ harmonic is associated to the mass only (indeed $\delta J_i=0$
in the Hessian eigenvector), while the $l=1$ harmonic is associated
to the angular momenta only (indeed $\delta M=0$ in the Hessian
eigenvector). In the singly-spinning black hole both the $l=0$ and
$l=1$ `harmonics' change the mass and the angular momentum.} This is
consistent with the fact that they are determined from the
thermodynamic Hessian \eqref{thermoHessian}, the changes in the
charges being proportional to the eigenvector $(\delta M, \delta
J_i)$. Notice that, being signalled by the Hessian of $S(M,J_i)$,
i.e. by the equation of state of the background black hole family,
the $l=0$ and $l=1$ zero-modes can only change the black hole into
another black hole of the {\it same} family, e.g. another MP black
hole.\footnote{In the present paper, these zero-modes give another
singly-spinning MP-AdS black hole. However, in the cohomogeneity-1
MP(-AdS) case \cite{Dias:2010eu}, the $l=1$ zero-mode breaks the
symmetries between the spins, as predicted by the eigenvector
$(\delta M, \delta J_i)$ having distinct $\delta J_i$. The zero-mode
therefore takes the black hole out of the equal spins sector, but
still in the general MP(-AdS) family.}

What is found by solving the problem of axisymmetric perturbations
is that  instabilities appear with new stationary zero-modes, not
predicted by the thermodynamic Hessian. We refer to them as {\it
non-thermodynamic} zero-modes. Their harmonic structure -- the
number of nodes, found to be $l\geq2$ -- is consistent with the
deformations of the event horizon proposed in
Refs.~\cite{Emparan:2007wm,Caldarelli:2008pz}; see
Figure~\ref{fig:phases}. These higher harmonics cannot change the
asymptotic charges, as exemplified explicitly in \cite{Dias:2010eu}.
The perturbed black holes will have the same mass and angular
momenta as the unperturbed solutions, and also, as imposed by the
boundary conditions, the same temperature and angular velocities.
Therefore, either they correspond to trivial gauge modes or to
bifurcations to new black hole solutions. We have shown in specific
cases \cite{Dias:2010eu,Dias:2010maa} that no regular gauge modes
are allowed. We conclude that the $l\geq2$ zero-modes mark a black
hole bifurcation, and also the onset of a classical ultraspinning
instability, one per each mode. That these linear modes do grow
exponentially with time was verified in \cite{Dias:2010eu}.

The ultraspinning conjecture of Ref.~\cite{Dias:2009iu} predicts
that zero-modes with $l\geq2$  can only occur for rotations higher
than that of the $l=1$ zero-mode. In the case of singly-spinning
MP-AdS black holes, this corresponds to the region $C$ of
Figure~\ref{fig:thermo6d}. This is based on the examples known, and
on the intuition that higher harmonics should be more stable than
lower harmonics. In general, when several distinct angular momenta
are present,  several different zero-modes with  $l=1$  may exist,
as exemplified in \cite{Dias:2010eu}. So the {\it ultraspinning
surface} (in our case, the surface between $B$ and $C$) is defined
as the surface in parameter space which encloses the region where
the thermodynamic Hessian \eqref{thermoHessian} has less than two
negative eigenvalues (in our case, regions $A$ and $B$).

Since this thermodynamic criterion is a necessary but not sufficient
condition,  to confirm the presence of these instabilities we still
have to solve the linear perturbations problem, and look for the
$l\geq2$ non-thermodynamic zero-modes. This study is done in
section~\ref{sec:PerturbationProblem}.

\subsection{Superradiant instability \label{subsec:superradiant}}

Before studying the ultraspinning instability of the singly-spinning
MP-AdS black hole,  we recall that this is not the only instability
present in the system. Indeed, it competes with the well-known
superradiant instability that afflicts rotating AdS black holes
whenever $\Omega_H\ell>1$
\cite{Cardoso:2004hs,Kunduri:2006qa,Cardoso:2006wa}. The
superradiant instability is associated to perturbations that break
the symmetry generated by $\partial_ \phi$, \ie it requires
perturbations whose angular momentum quantum number $m$ along the
$\phi$-direction is non-vanishing. The mechanism of the instability
is simple. A gravitational wave co-rotating with the background
black hole can extract rotational energy from the black hole if its
frequency $\omega$ satisfies the relation $\omega<m\Omega_H$, where
$\Omega_H$ is the angular velocity of the black hole. This
phenomenon is known as superradiance. On the other hand, the AdS
gravitational potential effectively acts as an asymptotically
reflecting box. The multiple superradiant amplifications and
reflections drive the system unstable if and only if
$\Omega_H\ell>1$.

Since there is a competition between the two instabilities, it is
relevant to  ask whether the ultraspinning instability occurs when
$\Omega_H \ell \le 1$. We find a negative answer. Indeed, in
Fig.~\ref{fig:thermo6d} we represent the superradiant surface,
$\Omega_H \ell= 1$, by a dashed line. It lays entirely in region
$B$. Black holes above this surface have $\Omega_H \ell > 1$, which
includes  all the  black holes in region $C$ that might also be
afflicted by the ultraspinning instability. Therefore,  all the  AdS
black holes that suffer from the ultraspinning instability that we
identify in the next section are also likely to be unstable under
superradiance.

\setcounter{equation}{0}
\section{Perturbations of MP-AdS black holes}
\label{sec:PerturbationProblem}

In this section, we will detail the linear perturbation problem,
which consists in studying the spectrum of the Lichnerowicz operator
on the black hole background. We will list the equations to be
solved and the boundary conditions, and we will briefly discuss
their numerical implementation.

\subsection{The Lichnerowicz eigenvalue problem} \label{subsec:ultraeigenvalue}

To find the onset of the ultraspinning instability we have to solve
the linearised Einstein  equations with a negative cosmological
constant. We search for stationary modes that depend on the radial
and polar coordinates, $r$ and $\theta$, but preserve the
$\mathbb{R}\times U(1)\times SO(d-3)$ symmetries of the background
MP-AdS black hole \eqref{mpbh}. The general ansatz for the perturbed
metric $h_{\mu\nu}$ satisfying these conditions
is:\footnote{Stationarity and axisymmetry lead to a metric
perturbation independent of $t$ and $\phi$, respectively. The
transverse $S^{d-4}$ line element is left unperturbed, apart from
its $(r,\theta)$-dependent scale factor, because we wish to preserve
the $SO(d-3)$ isometry. On the other hand, the components $tr$,
$t\theta$, $\phi r$, $\phi \theta$ can be discarded: the
corresponding TT gauge conditions and the linearised equations of
motion can be solved by separation of variables in $r$ and $\theta$,
and regularity implies that those components vanish.}
\begin{equation}
\begin{aligned}
ds^2 =& -\frac{\Delta_r(r)}{\Sigma(r,\theta)}\,e^{\delta\nu_0}\left[
dt-\frac{a\sin^2\theta}{\Xi} \,e^{\delta\omega}\,d\phi \right]^2 +
\frac{\sin^2\theta\,\Delta_\theta(\theta)}{\Sigma(r,\theta)}\,e^{\delta\nu_1}\left[
\frac{r^2+a^2}{\Xi}\,d\phi-a\, e^{-\delta\omega}\,dt\right]^2 \\
&
+\frac{\Sigma(r,\theta)}{\Delta_r(r)}\,e^{\delta\mu_0}\left[dr-\delta\chi\,
\sin\theta \,d\theta \right]^2+
\frac{\Sigma(r,\theta)}{\Delta_\theta(\theta)}\,e^{\delta\mu_1}\,d\theta^2
+r^2\cos^2\theta\,e^{\delta\Phi}\,
d\Omega^2_{(d-4)}\,,\label{ansatz}
\end{aligned}
\end{equation}
where $\{
\delta\nu_0,\delta\nu_1,\delta\mu_0,\delta\mu_1,\delta\omega,\delta\chi,\delta\Phi\}$
are   small quantities that describe our perturbations, and are
functions of  $(r,\theta)$ only. We will solve numerically the
coupled partial differential equations (PDEs) that govern these
perturbations. We choose to work in the traceless-transverse (TT)
gauge,
\begin{equation}
h^\mu_{\phantom{\mu}\mu}=0\, \qquad \hbox{and} \quad  \nabla^\mu
h_{\mu\nu}=0\,. \label{TT}
\end{equation}
The linearised Einstein-AdS equations are:
\begin{equation} (\tilde{\Delta}_L h)_{\mu\nu} \equiv - \nabla_\rho \nabla^\rho h_{\mu\nu}
-2\, R_{\mu\phantom{\rho}\nu}^{\phantom{\mu}\rho\phantom{\nu}\sigma}
h_{\rho\sigma}=0 \,, \label{LichnerowiczEq}
\end{equation}
where the operator $\tilde{\Delta}_L$ relates to the standard Lichnerowicz operator $\Delta_L$ as
\begin{equation} (\tilde{\Delta}_L h)_{\mu\nu} = (\Delta_L h)_{\mu\nu} + 2(d-1)\ell^{-2}h_{\mu\nu}\,. \label{LichnerowiczAdS}
\end{equation}
Following the strategy used to study the ultraspinning instability
in asymptotically flat  black holes in
\cite{Dias:2009iu,Dias:2010eu,Dias:2010maa}, we will actually
consider a more general eigenvalue problem,
\begin{equation} \label{eigenh}
(\tilde{\Delta}_L h)_{\mu\nu} =\lambda\, h_{\mu\nu}\,.
\end{equation}
More concretely, we will be looking for {\it negative modes} ($\lambda<0$) of the operator $\tilde{\Delta}_L$.

This eigenvalue problem arises in two instances. One is the
computation of quadratic quantum  corrections to the gravitational
partition function in the saddle point approximation
\cite{Gross:1982cv} (see \cite{Monteiro:2009ke} for the application
to the Kerr-AdS black hole). The quantum corrections present a
pathology whenever there is a negative mode. As we have mentioned,
each negative eigenvalue of the thermodynamic Hessian
\eqref{thermoHessian} gives a negative mode, which we refer to as
being {\it thermodynamic}.

In the asymptotically flat case, \eqref{eigenh} also  governs the
gravitational perturbations  of the form $e^{\ii \sqrt{-\lambda} z
}h_{\mu\nu}$ of the rotating black string that is constructed by
adding a flat extra dimension $z$ to the MP geometry. The AdS
analogue are the warped AdS black strings constructed in
Ref.~\cite{Park:2001jh}. These solutions are the generalisations of
the standard warped AdS black string, whose stability was studied in
\cite{Gregory:2000gf}, in the case where the transverse black hole
is also asymptotically AdS.\footnote{These should not be confused with the uniform black string solutions of Ref.~\cite{Brihaye:2007ju}, whose stability problem is not given by Eq. \eqref{eigenh}.}

Our strategy (following Refs.~\cite{Monteiro:2009ke,Dias:2009iu}) to
study the classical  perturbations ($\lambda=0$) of the black hole
is to look for a solution of \eqref{eigenh}, \ie a negative mode of
the black hole, and then vary the rotation parameter $a$ of the
black hole until the negative mode becomes a zero-mode, \ie until
$\lambda\rightarrow 0$. This strategy is motivated by the
availability of powerful numerical methods for solving eigenvalue
equations of the form \eqref{eigenh}. This strategy is motivated by
the availability of powerful numerical methods for solving
eigenvalue equations of the form (3.5). In particular, after an
adequate discretisation scheme using spectral methods
\cite{Trefethen}, (3.5) reduces to an algebraic eigenvalue problem
which we then solve using the inbuilt routine {\it Eingenvalues} of
Mathematica.

\subsection{Boundary conditions} \label{subsec:bcs}

To solve the Lichnerowicz eigenvalue system of equations
\eqref{eigenh} we have to  impose boundary conditions on the metric
perturbations \eqref{ansatz}. More concretely, we have to specify
boundary conditions at the horizon, $r=r_+$, at the asymptotic
infinity, $r\to\infty$, at the rotation axis $\theta=0$, and at the
equator $\theta=\pi/2$. The strategy to find the appropriate
boundary conditions was already discussed in great detail in section
4 of Ref.~\cite{Dias:2010maa}, where the ultraspinning instability
in asymptotically flat MP black holes was studied. The discussion
there translates straightforwardly to the present situation, and
hence we shall be brief.

We impose regularity of the metric perturbations on the event
horizon by demanding  regularity of the Euclideanised perturbed
geometry. The idea is to peform the standard Euclidean continuation
of the  black hole metric, and find the conditions required to have
regularity at the bolt (Euclideanised horizon). For the background
black hole, this demands that we identify
$(\tau,\phi)\sim(\tau,\phi+2\pi)\sim(\tau+\beta,\phi-\ii\,\Omega_H\beta)$,
where $\tau=\ii \,t$ is the Euclidean time, $\beta=1/T_H$ is the
inverse of the black hole temperature \eqref{MP:TempEnt}, and
$\Omega_H$ is its angular velocity \eqref{MP:OmegaH}. The boundary
conditions for the metric perturbations $h_{\mu\nu}$ can now be
determined by demanding  that $h_{\mu\nu} dx^\mu dx^\nu$ is a
regular symmetric 2-tensor on the background manifold. This requires
the following boundary conditions at the horizon, located at
$\rho\equiv[4(r-r_+)/\Delta_r'(r_+)]^{1/2}=0$,
\begin{equation}
\delta\chi,\,\delta\omega =O(\rho^2)\,, \qquad
\delta\nu_0-\delta\mu_0=O(\rho^2)\,,  \qquad
\partial_\rho\delta\mu_1,\,\partial_\rho\delta\nu_1,\,\partial_\rho\delta\Phi=O(\rho)\,.
\label{eqn:BC:r+}
\end{equation}
Note that we have not just imposed regularity of the perturbed
metric. If  $h_{\mu\nu} dx^\mu dx^\nu$ is a regular 2-tensor, then
perturbations obeying \eqref{eqn:BC:r+} preserve the angular
velocity and temperature of the background black hole. This is a
fundamental property to make the connection between the classical
ultraspinning instability and black hole thermodynamics that was
discussed in section \ref{sec:ultraspin} \cite{Dias:2010eu}.

To find the boundary conditions that the metric perturbations must
satisfy at the axis of rotation $(\theta=0)$, where $\partial_\phi$
vanishes, we proceed in the same way and require that $h_{\mu\nu}
dx^\mu dx^\nu$ is a regular symmetric 2-tensor, i.e. the components
$h_{\mu\nu}$ are regular when expressed in coordinates where the
background metric components are regular. Regularity as $\theta\to
0$ demands:
\begin{equation}
\delta\nu_1-\delta\mu_1=O(\theta^2)\,,\quad\partial_\theta
\delta\chi,\,\partial_\theta
\delta\omega,\,\partial_\theta\delta\mu_0,\,\partial_\theta\delta\nu_0,\,\partial_\theta
\delta\Phi  = O(\theta)\,. \label{eqn:BC:x1}
\end{equation}

Analogously, at the equator, $\theta= \pi/2$, we impose the boundary conditions:
\begin{equation}
\begin{aligned}
\delta\chi= O(x)\,, \quad
 \delta\Phi-\delta\mu_1=O(x^2)\,,\quad
 \partial_x\delta\omega,\,
 \partial_x\delta\mu_0,\,
 \partial_x\delta\nu_0,\,
 \partial_x\delta\nu_1=O(x)\,,
\end{aligned}
\label{eqn:BC:x0}
\end{equation}
as $x=\cos\theta \to 0$.

We have explicitly checked that the boundary conditions
\eqref{eqn:BC:r+},  \eqref{eqn:BC:x1} and \eqref{eqn:BC:x0} are
consistent both with the Lichnerowicz eigenvalue equations
\eqref{eigenh} and the TT gauge conditions \eqref{TT}. Indeed, the
first term in the series expansion of the eigenvalue equations
vanishes after we impose \eqref{eqn:BC:r+}, \eqref{eqn:BC:x1} and
\eqref{eqn:BC:x0}. On the other hand, we can use the TT gauge
conditions to express, \eg $\{ \delta\nu_0,\delta\nu_1,\delta\Phi\}$
as  functions of $\{
\delta\mu_0,\delta\mu_1,\delta\omega,\delta\chi\}$ and their first
derivatives. Again, the first term of a series expansion of these TT
gauge conditions is consistent with \eqref{eqn:BC:r+},
\eqref{eqn:BC:x1} and \eqref{eqn:BC:x0}.

At spatial infinity, $r\rightarrow\infty$, our second order
equations of motion allow  for two radial dependences of the metric
perturbation. We require that the boundary conditions preserve the
asymptotics of the background spacetime, in the sense that not only
(in our Boyer-Lindquist coordinates)
\begin{equation}
\label{eqn:BC:infinity}
h_{\mu\nu}{\bigr|}_{r\to \infty}\sim \frac
{1}{r^\alpha} \rightarrow 0 \,,
\end{equation}
for some constant $\alpha > 0$ that depends on the particular metric
component  and the number of spacetime dimensions $d$, but also in
the sense that the perturbations are normalisable. The negative
modes are ``gravitons with a positive mass'', and normalisability
requires that we pick the fastest decaying mode (cf. the
non-rotating case \cite{Prestidge:1999uq}). This happens to be the
mode that our numerical code can identify straightforwardly.

\subsection{Imposing the TT gauge conditions and the boundary conditions}
\label{sec:ttcond}

We wish to solve the Lichnerowicz eigenvalue problem \eqref{eigenh}
for the seven  metric perturbations described in \eqref{ansatz},
namely $\{
\delta\mu_0,\delta\mu_1,\delta\chi,\delta\omega,\delta\nu_0,\delta\nu_1,\delta\Phi\}$,
subject to the TT gauge conditions \eqref{TT}. The latter eliminate
three functions. We choose to solve the gauge conditions \eqref{TT}
for $\{ \delta\nu_0,\delta\nu_1,\delta\Phi\}$ in terms of $\{
\delta\mu_0,\delta\mu_1,\delta\chi,\delta\omega\}$ and their first
derivatives. Plugging this information in the full set of the
perturbation equations \eqref{eigenh}, we find that only four
equations remain of second order in $\{
\delta\mu_0,\delta\mu_1,\delta\chi,\delta\omega\}$. Explicitly,
these equations are:
\begin{equation}
\begin{aligned}
&(\tilde\triangle_L h)_{rr} =\lambda\, h_{rr}\,,\\
& (\tilde\triangle_L
h)_{r\theta} =\lambda\, h_{r\theta}\,,\\
& (\tilde\triangle_L h)_{\theta\theta} =\lambda\,
h_{\theta\theta}\,,\\
& a\,(\tilde\triangle_L h)_{tt}+\frac{\Xi \left( r^2+a^2+a^2\sin^2\theta
\right)}{\lp r^2+a^2\rp\sin^2\theta}\,(\tilde\triangle_L h)_{t\phi}
  + \frac{a \,\Xi^2}{\lp r^2+a^2\rp\sin^2\theta}\,(\tilde\triangle_L h)_{\phi\phi}\\
& \qquad\qquad\qquad\qquad
  =\lambda\, \left[ a\,h_{tt}+\frac{\Xi \left( r^2+a^2+a^2\sin^2\theta \right)}{\lp r^2+a^2\rp\sin^2\theta}\, h_{t\phi}
  + \frac{a \,\Xi^2}{\lp r^2+a^2\rp\sin^2\theta}\,h_{\phi\phi} \right]\,,
\end{aligned}
\label{eqn:finalsystem}
\end{equation}
and they describe the final set of equations that we have to solve.
A non-trivial  consistency check of our procedure is to verify that
the final  equations \eqref{eqn:finalsystem} imply that the
remaining equations in \eqref{eigenh} are automatically satisfied
(the latter equations are of third order in the independent
perturbation functions once the TT gauge conditions have been
imposed). We have explicitly verified that this is the case.

We will solve numerically the final eigenvalue problem
\eqref{eqn:finalsystem}   using spectral methods (see \eg
\cite{Trefethen}). To do so, we find it convenient to introduce new
radial and polar coordinates, \beq y=\frac{r}{r_+}-1\,,\qquad
x=\cos\theta\,, \eeq such that $0\leq y \leq \infty$ and $0\leq
x\leq 1$. The implementation of the method is simpler if all
functions obey Dirichlet boundary conditions on all the boundaries.
Therefore, we redefine our independent functions according to:
\begin{equation}
\label{eq:qs}
\begin{aligned}
&q_1(y,x) =  \left( \frac{r}{r_+}-1 \right) x(1-x)
\,\delta\mu_0(y,x)\,, \qquad \qquad
q_2(y,x) = r_m^{-1} (1-x)\,\delta\chi(y,x) \,, \\
&q_3(y,x) =  \left( \frac{r}{r_+}-1 \right)x(1-x)
\,\delta\mu_1(y,x)\,,  \qquad \qquad q_4(y,x) =  x
\,\delta\omega(y,x) \,,
\end{aligned}
\end{equation}
so that the $q_i$'s vanish at the boundaries. This guarantees that
the boundary  conditions  \eqref{eqn:BC:r+}, \eqref{eqn:BC:x1},
\eqref{eqn:BC:x0} and \eqref{eqn:BC:infinity} are correctly imposed.

In the limit $\ell\rightarrow \infty$, our equations and results
reduce to those  of the asymptotically flat case studied in
\cite{Dias:2009iu,Dias:2010maa}. In this case, we explicitly proved
that the ultraspinning zero-modes that we find cannot be pure gauge
modes. That is, there is no pure gauge perturbation, obeying the
boundary conditions we impose, which could potentially generate the
regular metric perturbations that we consider. By continuity, we
expect this property to hold true when the cosmological constant is
switched on.

\setcounter{equation}{0}
\section{Results and discussion}
\label{sec:results}

The problem of finding the onset of the axisymmetric ultraspinning
instability in  the singly-spinning MP-AdS black hole reduces, at
this point, to solving the equations \eqref{eqn:finalsystem} for the
metric perturbations \eqref{eq:qs} in the TT gauge and that obey
Dirichlet boundary conditions. The outcome of this analysis are the
dimensionless negative modes, $\lambda \,r_m^2$, of the modified
Lichnerowicz operator $\tilde{\Delta}_L$. These are obtained for
each pair of parameters that specify the black hole for a fixed AdS
radius $\ell$, namely the dimensionless mass radius, $r_m/\ell$, and
the dimensionless rotation parameter, $a/r_m$. We will later
describe the results in terms of the dimensionless asymptotic
charges $M/\ell^{d-3}$ and $J/\ell^{d-2}$, which are more physical
quantities. However, $r_m$ and $a$, which appear in the metric
\eqref{mpbh}, are more convenient for the numerical analysis.

\begin{figure}[t]
\centering
\includegraphics[width =7.0 cm]{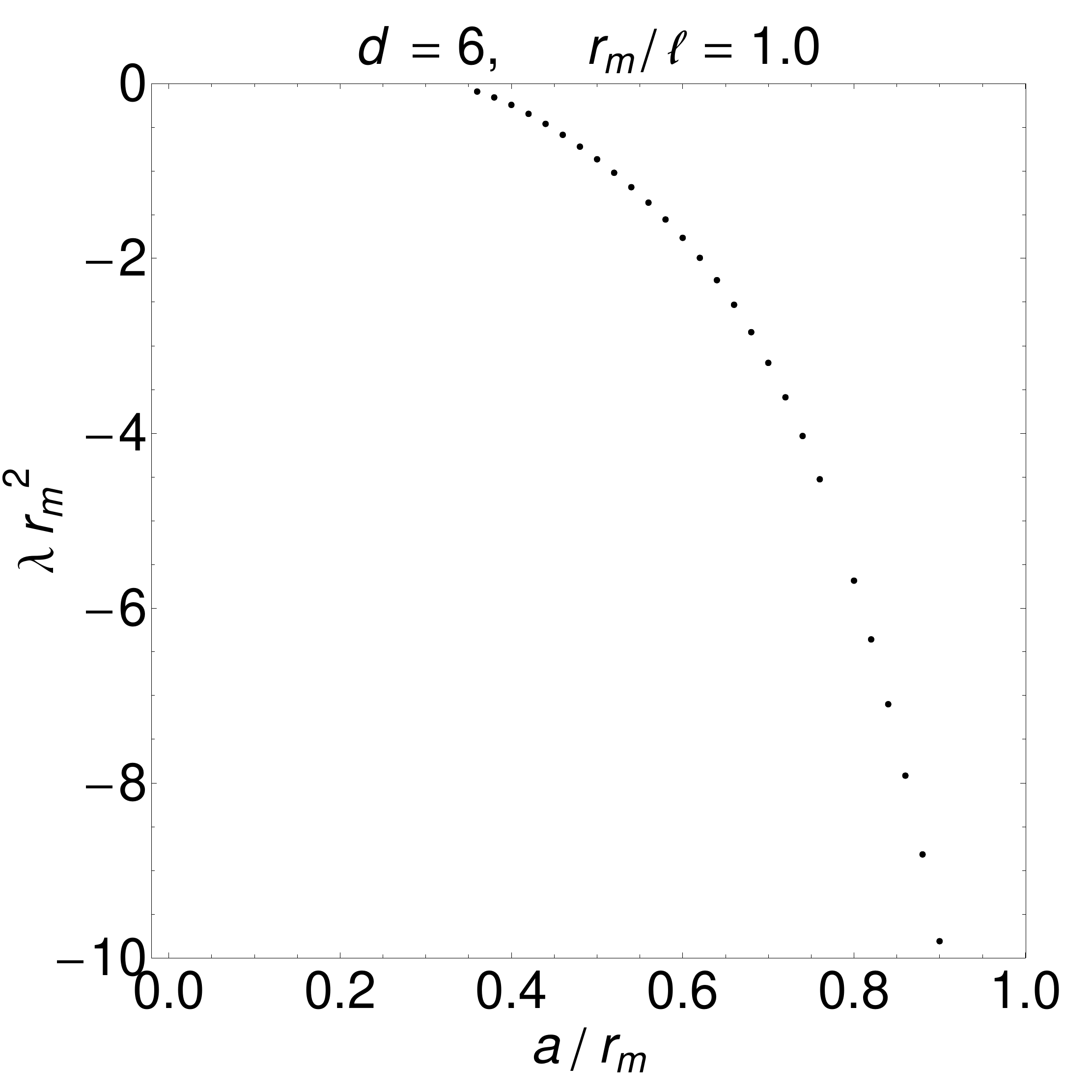}
\hspace{0.5cm}
\includegraphics[width =7.0 cm]{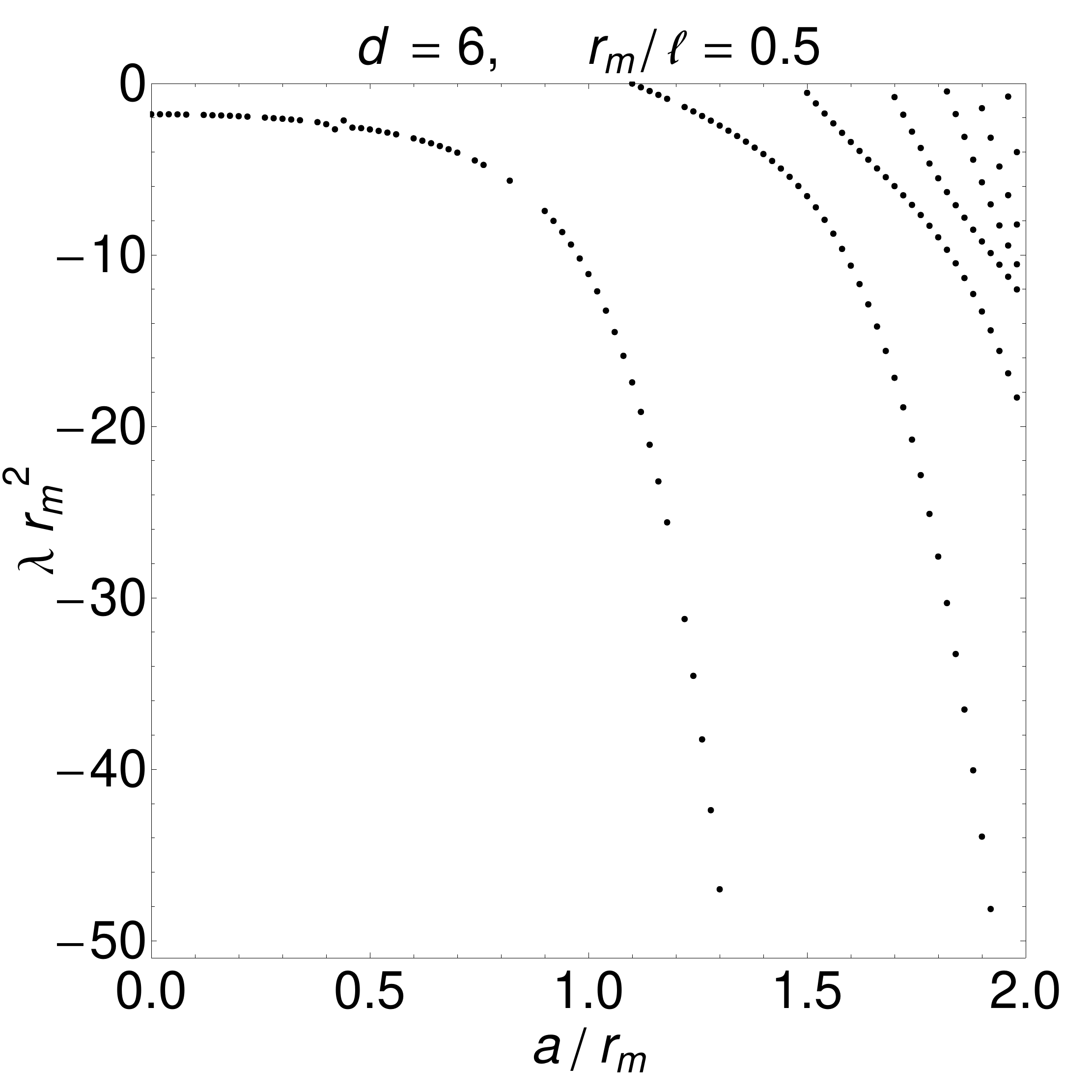}
\caption{\label{fig:negmodes} Dimensionless negative modes $\lambda r_m^2$ of the singly-spinning MP-AdS black hole in $d=6$ as a function of the dimensionless rotation parameter $a/r_m$ for fixed $r_m/\ell$. In the \textit{left} plot, we represent the spectrum for $r_m/\ell=1.0$, while the \textit{right} plot describes the spectrum for $r_m/\ell=0.5$. The latter has several (supposedly infinite) branches of zero-modes (for which $\lambda=0$) and the corresponding negative eigenvalues are labeled by the integer $l$. For this particular black hole family with $r_m/\ell=0.5$, the values of $a/r_m$ at which the first few branches intersect the $\lambda=0$ axis are:  $a/r_m{\bigl|}_{l=1}\simeq 1.10$, $a/r_m{\bigl|}_{l=2} \simeq 1.49$, $a/r_m{\bigl|}_{l=3} \simeq 1.69$, $a/r_m{\bigl|}_{l=4} \simeq 1.81$, and $a/r_m{\bigl|}_{l=5} \simeq 1.88$. As opposed to the thermodynamic $l=0,1$ zero-modes (curve on left plot, first two curves on right plot), the zero-modes with $l\geq 2$ describe the onset of ultraspinning instabilities of the black hole.}
\end{figure}

Examples of the spectrum of negative modes in $d=6$, as a function
of the  dimensionless rotation parameter, for two particular values
of $r_m/\ell$, are presented in Fig. \ref{fig:negmodes}. We expect
that for higher $d$ the spectrum is qualitatively similar to $d=6$
(we explicitly confirmed this for $d=7$).

The first property to notice is encoded in the curve on the left
plot and on its counterpart on the right plot, which is the first
curve counting from the left. This curve describes what we called
the thermodynamic $l=0$ negative mode in subsection
\ref{subsec:ultrathermo}. As discussed there, it was proven in
Ref.~\cite{Dias:2010eu} that this negative mode is always present if
the black hole is asymptotically flat. However, the AdS curvature
radius introduces a new scale in the system that changes this
property. For black holes with small $r_m/\ell$ (namely for
$r_m/\ell \lesssim 0.906$ in $d=6$) this negative mode is present
for any value of $a/r_m$, see \eg $r_m/\ell = 0.5$ in the right plot
of Fig.~\ref{fig:negmodes}, while for black holes with large
$r_m/\ell$ (namely for $r_m/\ell \gtrsim 1.587$ in $d=6$) the
thermodynamic $l=0$ eigenvalue is positive, independently of the
rotation. When we fix $r_m/\ell$ to be in between these two regimes,
the thermodynamic $l=0$ eigenvalue is positive for small $a/r_m$ and
becomes negative for large $a/r_m$, see \eg the $r_m/\ell = 1.0$
case in the left plot of Fig.~\ref{fig:negmodes}. The phase
transition, where the thermodynamic $l=0$ eigenvalue is a zero-mode,
occurs precisely at the critical rotation identified in
\eqref{Hessian0modeL0}, namely at $a/r_m{\bigl|}_{l=0} \simeq 0.337$
(for  $r_m/\ell = 1.0$ and in $d=6$). This is what we expect from
the thermodynamics discussion of subsection
\ref{subsec:ultrathermo}: in regions $B,C$ of
Fig.~\ref{fig:thermo6d} the thermodynamic $l=0$ eigenvalue is
negative while in region $A$ it is positive. The agreement between
the numerical and thermodynamic results is an important check of the
numerical code.

The next interesting property of the spectrum of negative modes is
encoded in  the second curve counting from the left on the right
plot of Fig.~\ref{fig:negmodes}. According to
Fig.~\ref{fig:thermo6d}, black holes with $r_m/\ell \lesssim 0.864$
in $d=6$ have a negative (positive) thermodynamic $l=1$ eigenvalue
if the rotation parameter is such that the black hole is in region
$C$ (region $B$), with the $l=1$ zero-mode curve defined by
\eqref{Hessian0modeL1}. The black hole family displayed on the right
plot of Fig.~\ref{fig:negmodes} obeys these conditions, since
$r_m/\ell = 0.5$. Indeed, for rotations below (above) the critical
rotation $a/r_m{\bigl|}_{l=1}  \simeq 1.10$, the thermodynamic $l=1$
eigenvalue is positive (negative) as predicted by
\eqref{Hessian0modeL1}.

The $l=0$ and $l=1$ negative eigenvalue curves are the only ones
whose existence can  be predicted by the analytical thermodynamic
analysis of subsection \ref{subsec:ultrathermo}. As discussed there,
Refs.~\cite{Reall:2001ag,Monteiro:2009ke,Dias:2010eu} proved that a
zero-mode of the thermodynamic Hessian \eqref{thermoHessian} is also
a zero-mode of the (modified) Lichnerowicz operator \eqref{eigenh},
but the latter can have additional zero-modes, unrelated to that
Hessian. The thermodynamic instabilities associated with the $l=0,1$
negative modes do not correspond to classical gravitational
instabilities of the black hole. In particular, the thermodynamic
$l=0,1$ zero-modes do not describe the onset of a classical
instability. Instead, they describe gravitational perturbations that
change the mass and the angular momentum of the black hole, within
the MP-AdS family, but preserve the temperature and the angular
velocity, due to the boundary conditions.

The thermodynamic $l=1$ zero-mode curve defines the ultraspinning
surface of the system.  It was conjectured in
Ref.~\cite{Dias:2009iu} that only black holes which are outside this
surface (\ie in region $C$ of Fig.~\ref{fig:thermo6d}) might develop
new zero/negative modes that have no thermodynamic interpretation,
and that describe classical ultraspinning instabilities which are
axisymmetric. The right plot of Fig.~\ref{fig:negmodes} confirms
that the spectrum of MP-AdS black holes indeed has (a supposedly
infinite sequence of) new branches of non-thermodynamic zeros-modes,
for rotations strictly higher than the critical rotation
$a/r_m{\bigl|}_{l=1}$ associated with the $l=1$ zero-mode. These
families exhibit an underlying harmonic structure (although the
equations that we solve do not seem to separate into radial and
angular equations). We use this property to suggestively label the
several new branches by successive integers $l=2,3,4,\ldots$. This
is an appropriate notation since the integer $l$ coincides with the
number of nodes that the metric perturbations have on the horizon
$y=0$ \footnote{For the particular case where $\ell\rightarrow
\infty$, this statement is illustrated in Fig.~3 of
\cite{Dias:2010maa}, where we plot the functions $\delta\chi(x,y)$
and $\delta\omega(x,y)$ for $l=1$ and $l=2$. The figures are
qualitatively similar for finite $\ell$.\label{footnotepinches}}.
The values of $a/r_m$ at which the first few branches intersect
$\lambda=0$ for the particular family of black holes with
$r_m/\ell=0.5$ in $d=6$ are summarised in the caption of
Fig.~\ref{fig:negmodes}. The spectrum of black hole families with
$r_m/\ell \lesssim 0.906$ is qualitatively similar to this
particular case.

\begin{figure}[t]
\centering
\includegraphics[width =7.0 cm]{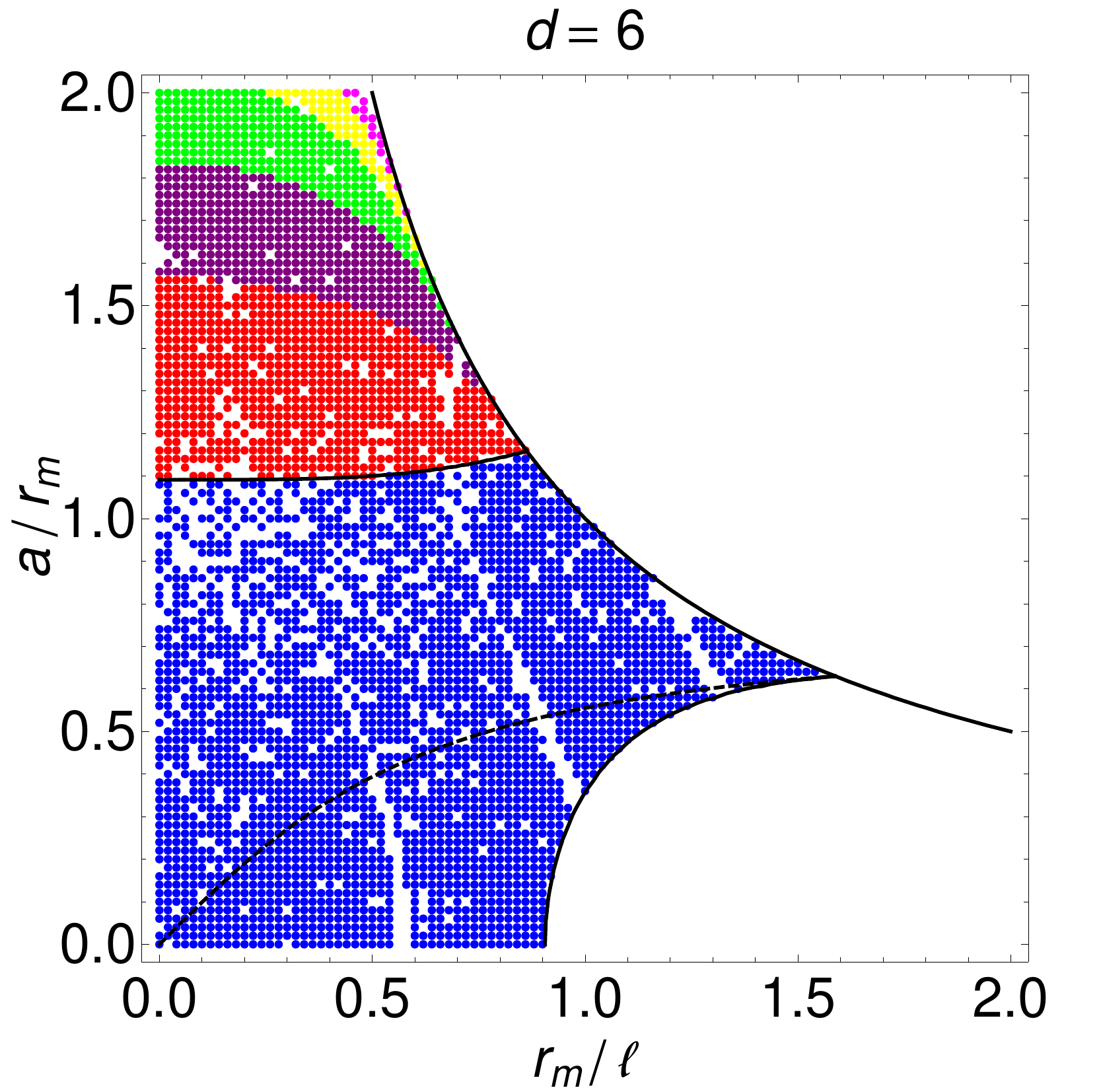}
\hspace{0.5cm}
\includegraphics[width =7.0 cm]{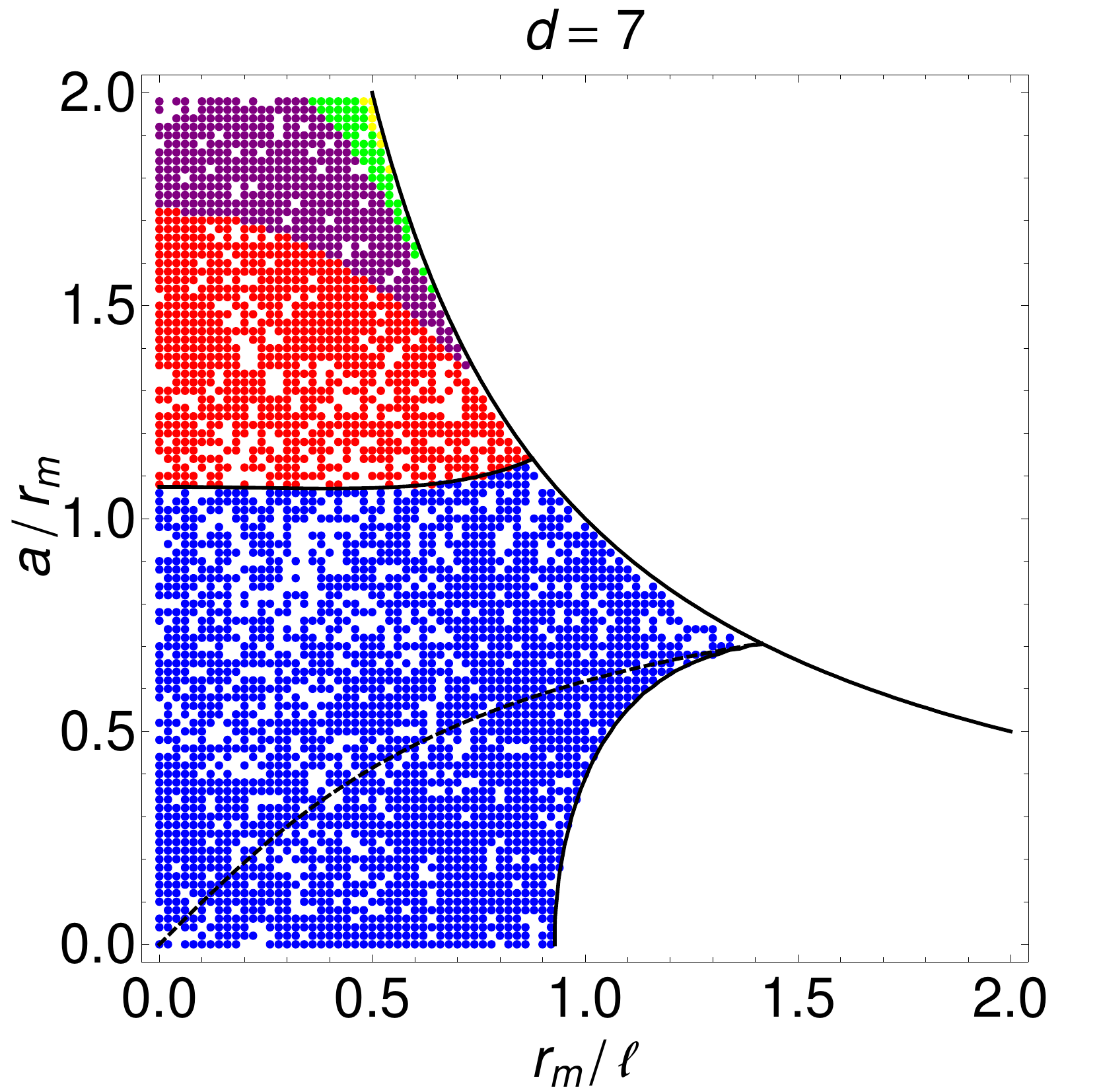}
\caption{\label{fig:spectrum} Number of negative modes of the
singly-spinning MP-AdS  black hole in $d=6$ \textit{(left)} and
$d=7$ \textit{(right)}. The plots describe the dimensionless
rotation parameter $a/r_m$ as a function of the dimensionless
mass-radius $r_m/\ell$. As we move from the bottom to the top, the
new colored/dotted areas represent regions where a new negative mode
of the (modified) Lichnerowicz operator gets excited: blue (one
negative mode), red (2 n.m.), purple (3 n.m.), green (4 n.m.),
yellow (5 n.m.), pink (6 n.m.), \ldots\, The interpretation of the
several curves plotted is described in Fig.~\ref{fig:thermo6d}.
(Note that the faults in these figures, e.g. where we expected to
find blue dots, correspond to parameter space points where our
numerical code failed. They do {\it not} correspond to black holes
with no negative modes.)}
\end{figure}

In Fig.~\ref{fig:spectrum}, we plot the number of negative modes
detected numerically,  for a grid of points in the parameter plane
$\lp r_m/\ell,a/r_m\rp$ in $d=6$ (left plot) and in $d=7$ (right
plot). This figure completes the information displayed in
Fig.~\ref{fig:thermo6d} (in particular, for $d=6$, the black curves
are the same plotted in Fig.~\ref{fig:thermo6d}). Not only it
includes the numerical and analytical results for the thermodynamic
$l=0,1$ modes but, in addition, it adds the numerical results
relative to the $l\geq 2$ modes that have no thermodynamic
interpretation. In region $A$ (see Fig.~\ref{fig:thermo6d}), we find
no negative modes. In region $B$, the system has only the $l=0$
negative mode (blue dotted region in Fig.~\ref{fig:spectrum}). In
region $C$, the $l=1$ negative mode is also present, but additional
negative modes appear for higher rotations. In
Fig.~\ref{fig:spectrum}, these are signaled by the red, purple,
green, yellow and pink dotted areas. In the red area, only the
$l=0,1$ modes are negative, and there is no classical instability.
However, for higher rotation, in the purple area and above it, the
$l=2$ harmonic becomes another negative mode of the modified
Lichnerowicz operator \eqref{eigenh}. The transition from the red
into the purple area defines the {\it non-thermodynamic $l=2$
zero-mode curve}. This surface signals the onset of the
ultraspinning instability in the parameter space. The boundary of
the purple and green areas identifies the {\it non-thermodynamic
$l=3$ zero-mode curve}: above it the $l=3$ harmonic is also excited,
\ie it becomes a negative mode. Similarly, the green/yellow
(yellow/pink) transition marks the {\it non-thermodynamic $l=4$
($l=5$) zero-mode curve} above which the $l=4$ ($l=5$) harmonic is
also a negative mode. Not shown in this figure, there should be an
infinite sequence of new excited harmonics. These would become
visible if we considered higher values of $a/r_m$ in region C.

\begin{figure}[t]
\centering
\includegraphics[width =8.0 cm]{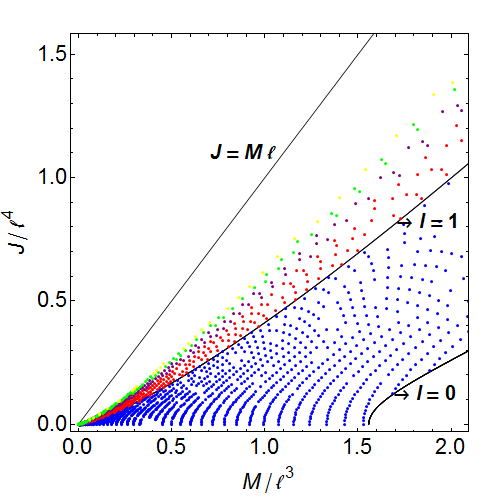}
\caption{\label{fig:modesMvsJ} Number of negative modes of the
singly-spinning MP-AdS black hole in $d=6$.  This figure has the
data of Fig.~\ref{fig:spectrum}-left, now plotting the dimensionless
angular momentum $J/\ell^{d-2}$ as a function of the dimensionless
mass $M/\ell^{d-3}$. Regular MP-AdS black holes exist for
$|J|<M\ell$. We represent the location of the $l=0$ and the $l=1$
zero-modes, for reference.}
\end{figure}

In Fig.~\ref{fig:modesMvsJ}, we proceed to a thorough analysis of
the singly-spinning  MP-AdS parameter space. We represent the data
in Fig.~\ref{fig:spectrum} for $d=6$ in terms of the dimensionless
asymptotic conserved charges, $M\ell^{d-3}$ and $J\ell^{d-2}$,
instead of using $r_m$ and $a$ to specify the black hole.
We do this for the sake of clarity, because the charges are more
physical,  and also because the map $(r_m,a) \to (M,J)$ in
Eqs.~\eqref{MP:MJ} is singular as $|a|\to \ell$, so that the
parameter space looks very different. The physical parameter range
is strictly below the curve $|J|=M\ell$. To compare this figure with
Fig.~\ref{fig:spectrum}-left, notice that: (i) the asymptotically
flat limit, which was the $r_m/\ell =0$ line in
Fig.~\ref{fig:spectrum}-left, is the point $(0,0)$ in
Fig.~\ref{fig:modesMvsJ}; (ii) the line $a=\ell$, for finite
$r_m/\ell$, in Fig.~\ref{fig:spectrum}-left is now the asymptotic
corner given by the limit $R \to \infty$ of $(R,R)$ in
Fig.~\ref{fig:modesMvsJ}, so that the various coloured regions do
not all meet in the parameter space as suggested in
Fig.~\ref{fig:spectrum}-left; and (iii) the limit $a/r_m \to
\infty$, within $a<\ell$, in Fig.~\ref{fig:spectrum}-left is the
curve $J=M\ell$ that puts a bound on the parameter space in
Fig.~\ref{fig:modesMvsJ}. Notice that, for relatively small fixed
values of the mass (in AdS units), say $M/\ell^{d-3} = 1$, there is
always an $l=0$ negative mode, as in the asymptotically flat limit.
On the other hand, there is a $l=0$ zero-mode for larger masses, say
$M/\ell^{d-3} = 2$. For any fixed value of the mass, we find that a
sufficiently high rotation leads to the ultraspinning instabilities,
in correspondence with Fig.~\ref{fig:phases}.

Our study is limited to \textit{stationary} perturbations.
Therefore, we have  not proven that our $l\geq 2$ negative modes do
indeed correspond to a region in parameter space where some black
hole perturbations grow exponentially with time. Including
time-dependence is not conceptually difficult but it is only
computationally harder. Additional metric components would have to
be perturbed (\eg $h_{tr}$, \ldots) in \eqref{ansatz}, and we would
have to solve numerically a system of many more PDEs. This problem
is however of fundamental interest since its solution would provide
a definite proof of the ultraspinning instability in MP-AdS black
holes, together with information on the instability timescale.
Having said this, we nevertheless claim that we have found the
stationary zero-modes that signal the onset of the ultraspinning
instability. Our confidence comes from the fact that the analogous
stability problem, including time-dependence, was studied in
\cite{Dias:2010eu} for asymptotically flat cohomogeneity-1 MP black
holes (equal angular momenta, odd $d$). There, the analogues of the
$l=1,2,3,4,\ldots$ zero-modes are also present, and the
corresponding perturbation sectors decouple. The time-dependent
analysis confirmed the absence of a classical black hole instability
in the $l=1$ sector of perturbations, and the existence of an
ultraspinning instability in the $l\geq2$ sectors. We take this
result as good evidence in support of a similar interpretation in
the singly-spinning MP-AdS case.

The ultraspinning $l\geq 2$ zero-modes do not admit a thermodynamic
interpretation,  as we have stressed. They should correspond not
only to the onset of classical instabilities, but also to the
bifurcation to new stationary AdS black hole families with `pinched'
horizons, which have the same isometries as the MP-AdS black holes.
This is the conjecture of Ref.~\cite{Caldarelli:2008pz} (following
\cite{Emparan:2003sy,Emparan:2007wm} in the asymptotically flat
case), represented in Fig.~\ref{fig:phases}. In particular, the
$l=2$ zero-mode should connect to a family of AdS black holes which
are `pinched' on the poles. The location of the $l=2$ zero-mode is
represented as point $I$ in Fig.~\ref{fig:phases}. At some point
along the phase diagram of this new family, a horizon topology phase
transition should connect it to the AdS black ring with the same
isometries, which has been constructed perturbatively in the limit
of large dimensionless angular momentum \cite{Caldarelli:2008pz}.
Similarly, the $l=3$ ($l=4$) zero-mode, identified as point $II$
($III$) in Fig.~\ref{fig:phases}, should connect to a new family
that interpolates between the MP-AdS family and the AdS black Saturn
(concentric rings), constructed perturbatively in
\cite{Caldarelli:2008pz}. Our results support the conjecture of
Ref.~\cite{Caldarelli:2008pz} not only because we find the
bifurcation points, but also because the shape of the numerical
perturbations, with its harmonic structure labelled by $l$, is
consistent with the appropriate `pinches' of the horizon (footnote
\ref{footnotepinches}).

As discussed in subsection~\ref{subsec:superradiant}, all the
ultraspinning unstable MP-AdS  black holes in the purple, green,
yellow, pink, \ldots\, dotted areas of Fig.~\ref{fig:spectrum} have
$\Omega_H\ell>1$, and thus all of them are also afflicted by the
superradiant instability
\cite{Cardoso:2004hs,Kunduri:2006qa,Cardoso:2006wa}. The instability
is expected to be inherited by the `pinched' black hole families, at
least close to the bifurcation point.

An interesting open problem is what is the interpretation of the
ultraspinning instability  and of the associated new AdS pinched
black hole phases  in the holographic dual field theory. Steps in
this direction have already been taken in the context of the
Scherk-Schwarz compactified AdS. In this case, in the hydrodynamic
limit of the holographic theory one finds: i) pinched plasma balls
\cite{Lahiri:2007ae} that should be in correspondence with the
gravitational ultraspinning system, ii) deformed plasma tubes
\cite{Cardoso:2006ks} that should be dual to the Gregory-Laflamme
instability of Scherk-Schwarz-AdS black strings and iii) rotating
plasma ball instabilities \cite{Cardoso:2009bv} that should describe
the gravitational bar-mode instability
\cite{Shibata:2009ad,Shibata:2010wz} mentioned in the Introduction.

\section*{Acknowledgements}

It is a pleasure to acknowledge the stimulating discussions with our collaborators Roberto Emparan  and Harvey Reall in the ultraspinning project. OJCD acknowledges financial support provided by the European Community through the Intra-European Marie Curie contract PIEF-GA-2008-220197. This work was partially funded by FCT-Portugal through projects PTDC/FIS/099293/2008, CERN/FP/83508/2008 and CERN/FP/109306/2009. PF was a member of the CPT group at the Department of Mathematical Sciences, Durham University, and was supported by an STFC rolling grant while part of this work was carried out. PF is supported by an EPSRC postdoctoral fellowship [EP/H027106/1].


\end{document}